%% file: Attention_MIL.tex
\documentclass[journal]{IEEEtran}
\usepackage{cite}
\ifCLASSINFOpdf
  \usepackage[pdftex]{graphicx}
\else
  \usepackage[dvips]{graphicx}
  \DeclareGraphicsExtensions{.eps}
  \graphicspath{{./figures/}}
\fi
\usepackage{amsmath}
\usepackage{amssymb}

\usepackage{array}
\usepackage{makecell}
\usepackage{booktabs}

\ifCLASSOPTIONcompsoc
 \usepackage[caption=false,font=normalsize,labelfont=sf,textfont=sf]{subfig}
\else
 \usepackage[caption=false,font=footnotesize]{subfig}
\fi

\usepackage{dblfloatfix}

\begin{document}
\bstctlcite{IEEEexample:BSTcontrol}
    \title{A Multi-resolution Model for Histopathology Image Classification and Localization with Multiple Instance Learning}
  \author{Jiayun~Li,~\IEEEmembership{} 
    Wenyuan~Li,~\IEEEmembership{}
    Anthony~Sisk,~\IEEEmembership{} 
    Huihui~Ye,~\IEEEmembership{} W.~Dean~Wallace,~\IEEEmembership{}
    William~Speier,~\IEEEmembership{}
	and~Corey~W.~Arnold~\IEEEmembership{}

  \thanks{This work was supported in part by NIH/NCI R21CA220352. (Corresponding author: Corey W. Arnold.)}
  \thanks{J. Li, W. Li, W. Speier and C.W. Arnold are from the Computational Diagnostics Lab, and the Departments of Radiological Sciences and Pathology \& Laboratory Medicine, UCLA, 924 Westwood Blvd, Suite 420, Los Angeles, CA 90024 USA (e-mail: jiayunli@ucla.edu, cwarnold@ucla.edu).}%
  \thanks{A. Sisk, H. Ye are from the Departments of Radiological Sciences and Pathology and Laboratory Medicine, UCLA, 924 Westwood Blvd, Suite 420, Los Angeles, CA 90024 USA}
  \thanks{W. D. Wallace is from the Department of Pathology, USC, 2011 Zonal Avenue, Los Angeles, CA 90033 USA}
  \thanks{Colour figures are available online.}
  \thanks{Copyright (c) 2017 IEEE. Personal use of this material is permitted. However, permission to use this material for any other purposes must be obtained from the IEEE by sending a request to pubs-permissions@ieee.org.}
}

\markboth{IEEE Journal of Biomedical and Health Informatics}{Li \MakeLowercase{\textit{et al.}}: A Multi-resolution Model for Histopathology Image Classification and Localization with Multiple Instance Learning}

\maketitle

\begin{abstract}
 \input{./tex/abstract}
\end{abstract}

\begin{IEEEkeywords}
Whole slide image, multiple instance learning, prostate cancer, attention, multiple resolution, cancer grade prediction 
\end{IEEEkeywords}

%
\IEEEpeerreviewmaketitle

\section{Introduction}
\input{tex/introduction.tex}

\section{Method}
\input{tex/method.tex}

\section{Experiment}
\input{tex/experiment.tex}

\section{Results}
\input{tex/result.tex}

\section{Discussion}
\input{tex/discussion.tex}

\section{Limitations and Future Work}
\input{tex/limitations.tex}

\section{Conclusion}
\input{tex/conclusion.tex}

\ifCLASSOPTIONcaptionsoff
  \newpage
\fi

\bibliographystyle{IEEEtran}
\bibliography{IEEEabrv,Bibliography}


\end{document}

%% file: tex/abstract.tex
Histopathological images provide rich information for disease diagnosis. Large numbers of histopathological images have been digitized into high resolution whole slide images, opening opportunities in developing computational image analysis tools to reduce pathologists' workload and potentially improve inter- and intra- observer agreement. Most previous work on whole slide image analysis has focused on classification or segmentation of small pre-selected regions-of-interest, which requires fine-grained annotation and is non-trivial to extend for large-scale whole slide analysis. In this paper, we proposed a multi-resolution multiple instance learning model that leverages saliency maps to detect suspicious regions for fine-grained grade prediction. Instead of relying on expensive region- or pixel-level annotations, our model can be trained end-to-end with only slide-level labels. The model is developed on a large-scale prostate biopsy dataset containing 20,229 slides from 830 patients. The model achieved 92.7\% accuracy, 81.8\% Cohen's Kappa for benign, low grade (\emph{i.e.} Grade group 1) and high grade (\emph{i.e.} Grade group $\geq$ 2) prediction, an area under the receiver operating characteristic curve (AUROC) of 98.2\% and an average precision (AP) of 97.4 \% for differentiating malignant and benign slides. The model obtained an AUROC of 99.4\% and an AP of 99.8\% for cancer detection on an external dataset. 

%% file: tex/introduction.tex
\label{sec:introduction}

\IEEEPARstart{P}{rostate} cancer accounts for nearly 20\% of new cancer diagnosed in men, and is the most prevalent and second deadliest cancer in men in the United States\cite{siegel2019cancer}. Active surveillance (AS) is an important management option for patients with clinically localized low- to intermediate-risk prostate cancer \cite{tosoian2016active}. Prostate biopsy, which plays an essential role in treatment planning, is performed repeatedly during the course of the AS. Each biopsy can result in several tissue slides that are examined and, if cancer is present, assigned Gleason scores (GS) by pathologists based on the Gleason grading system. The GS is determined by two most predominant Gleason patterns that range from 1 (G1), closely resembling normal glands and carrying the lowest risk for dissemination, to 5 (G5), representing undifferentiated carcinoma and exhibiting the highest risk for dissemination. A recent study proposed to revise the Gleason grading system with 5 Gleason Grade groups (GGs) to reduce the over-treatment of low-grade prostate cancer \cite{epstein2016contemporary}: GG 1 (GS $\leq$ 6), GG 2 (G3 + G4), GG 3 (G4 + G3), GG 4 (GS = 8) and GG 5 (GS $\geq$ 9). Patients with intermediate- to high-risk localized prostate cancer (GG $\geq$ 2) may be intervened with radiotherapy and radical prostatectomy, with or without hormonal therapy. 

Currently, the diagnosis of prostate cancer relies on pathologists to examine multiple levels of biopsy cores at the scanning magnification, and identify suspicious regions for high power examination and immunohistochemistry if necessary. This process can be tedious and time-consuming. More importantly, some patterns, e.g. ill-defined G4 versus tangentially sectioned G3, are prone to inter- and intra-observer variability. Therefore, the current clinical practice can be improved by computer aided diagnosis tools (CAD) that can function as primary screening, to localize suspicious regions, and be utilized as a second reader for Gleason grading. Deep learning-based CAD models have been developed and demonstrated promising performance in many medical imaging fields \cite{shen2017deep, thrall2018artificial, currie2019machine, chaddad2018multimodal, chaddad2018predicting}. However, the enormous size of whole slide images (WSI), the variability in tissue appearances, and the artifacts incurred during staining and scanning impose many unique challenges in developing such CAD tools.

\subsection{Related Work}
Classification of small homogeneous regions of interest (ROIs) pre-selected by pathologists has been the main focus of most early work in WSI image analysis \cite{farjam2007image, doyle2010boosted, nguyen2012prostate}. Though these methods have achieved good results, they cannot be easily extended to handle regions with heterogeneous tissue types because they require a set of manually selected tiles with the same cancer grade, which is non-trivial to obtain.    
Some work has addressed this challenge by developing segmentation models that can provide pixel-wise predictions for tiles with various tissue contents \cite{gertych2015machine,li2018based, li2017multi, ing2018semantic, li2018path}. However, these models still analyzed tiles instead of the entire slide.   

With an increasing number of scanned slides and computing power, research in WSI has been shifting to slide-level analysis \cite{nagpal2019development, nagpal2020development, nir2019comparison}. For example, in a recent work by Nagpal \emph{et-al.} \cite{nagpal2020development}, they developed a two-stage model on a dataset containing 752 biopsy slides and achieved 71.7\% Cohen’s Kappa for predicting benign, GG1, GG2, GG3, and GG4-5. However, the model relied on a large amount of expensive fine-grained annotations. 

While these papers demonstrated promising performance in slide-level predictions \cite{nagpal2019development,litjens2016deep, nagpal2020development}, they required a large number of expensive pixel or patch-level manual annotations for training. Bulten \emph{et-al.} utilized a semi-automated labeling technique for prostate biopsy slide classification \cite{bulten2019automated, bulten2020automated}. Specifically, the authors used a pre-trained tissue segmentation network to identify tissue areas, within which cancerous regions were localized by a pre-trained tumor detection network. Non-epithelial areas were excluded from identified cancerous regions with an epithelium detection model. Detected epithelial areas from slides with a single Gleason pattern inherited slide-level labels and formed their initial training set for a U-Net model. Slide-level predictions were determined by percentage of Gleason patterns obtained from the segmentation network. However, this framework was built upon three pre-trained preprocessing modules, each of which still required pixel-wise annotations.  

Multiple instance leaning (MIL) \cite{andrews2003support, dietterich1997solving} has been utilized to address weakly-supervised learning challenges in tumor detection \cite{melendez2015combining, melendez2014novel, quellec2016multiple}, segmentation \cite{jia2017constrained, xu2017large}, and classification \cite{hou2016patch, tennakoon2019classification, mercan2017multi, wang2019rmdl, ilse2018attention, yan2016multi}. Most MIL models fall roughly into two general categories \cite{amores2013multiple,cheplygina2019not, carbonneau2018multiple}: instance-base and bag-based methods. 
Bag-based methods usually demonstrate better performance for tasks where global (\emph{i.e.,} bag-level) predictions are more important. Nevertheless, they suffer from a lack of interpretability, since instance predictions are often unavailable \cite{ilse2018attention}. Ilse \emph{et-al.} developed an attention-based MIL model that can visualize the relative contribution of instances for final prediction through a trainable attention module without sacrificing bag-level prediction performances \cite{ilse2018attention}. The model was utilized to identify epithelial and malignant patches within small tiles extracted from WSI for colon cancer and breast cancer datasets, respectively. However, they did not address the challenge of classifying much larger and more heterogeneous WSIs. Moreover, they only utilized attention maps for visualization.  

Few recent works have utilized MIL for whole slide classification \cite{wang2019rmdl,wang2020ud, campanella2019clinical, lu2020data}. Campanella \emph{et-al.} employed an instance-based approach to discriminate between malignant and benign prostate WSIs \cite{campanella2019clinical}. 
They considered the top \textit{k} tiles with the highest probabilities from positive slides after applying the CNN model as pseudo positive training samples, which were updated in each training epoch. 
In the second stage, they investigated aggregation functions to produce a final slide-level prediction. The model achieved promising performance on three different types of large-scale clinical datasets. However, the more difficult problem of Gleason grading was not investigated in the paper.

In this paper, we proposed a multi-resolution MIL-based (MRMIL) model for prostate biopsy WSI classification and weakly-supervised tumor region detection. Different from most existing studies, which rely on highly curated datasets with fine-grained manual annotations at pixel- or region-level, our model can be trained with only slide-level labels obtained from pathology reports. Similar to how WSIs are typically reviewed by pathologists, the proposed model scans through the entire slide to localize suspicious regions at a lower resolution (\emph{i.e.}, at 5x), and then zooms in on the suspicious regions to make grade predictions (\emph{i.e.}, at 10x). The main contribution of this paper can be summarized as the following:

1) We developed a novel MRMIL model, which can be trained with only slide-level labels, for prostate cancer WSI classification and detection.  
2) We trained and validated our model on a large dataset, containing 13,145 slides from 661 patients, which were retrieved from clinical cases without manual curation. To have a better understanding of our model's performance, we also visualized the data representations learned by the model.  
3) We tested on an independent test set consisting of 7,114 biopsy slides from 169 patients and an external dataset. The model achieved 81.8\% Cohen's Kappa ($\kappa$) and 92.7\% accuracy (Acc) for classifying benign, low grade (\emph{i.e.}, GG = 1), and high grade (\emph{i.e.}, GG $\geq$ 2) slides. Additionally, We extended our best model for Gleason group prediction, and it obtained 71.1\% $\kappa$ and 86.8\% quadratic $\kappa$. 


%% file: tex/method.tex
\label{sec:methods}
\begin{figure*}[!tbh]
\centering
\includegraphics[width=4.8in]{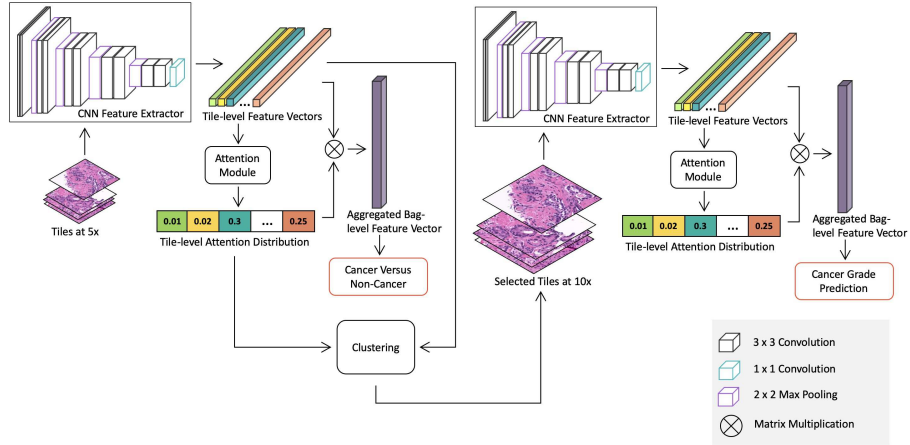}
\caption{Overview of the proposed whole slide image detection and classification model. The model consists of two stages: a cancer detection stage at a low magnification and a cancer classification stage at a higher magnification for suspicious regions. Both stages contains a CNN feature extractor, which is trained in the MIL framework with slide-level labels. Specifically, the detection stage model is trained with all tiles extracted from slides at 5x to differentiate between benign and malignant slides. The attention module in the detection stage model produces a saliency map, which represents relative importance of each tile for predicting slide-level labels. Then we use the K-means clustering method to group tiles into clusters based on tile-level features. The number of tiles selected from each cluster is determined by the mean of cluster attention values. Discriminative tiles identified by the detection stage model are then extracted at 10x and fed into the classification stage model for cancer grade classification.}

\label{model_overview}
\end{figure*}
\subsection{Problem Definition}
Due to the enormous size of WSIs, slides are usually divided into smaller tiles for analysis. However, different from works that utilized fine-grained manual annotations, our model is developed on the dataest with only slide-level labels (\emph{i.e.,} We don't have labels for each tile, instead, we only have a slide-level label for a set of tiles.). Therefore, we formulate the WSI classification problem in the MIL framework. Specifically, a slide is considered as one bag. $k$ tiles of size $N \times N$ extracted from the bag are denoted as instances within the bag, each of which may have different instance-level labels $y_{i}, i \in [1, k]$. During training, only the label for a set of instances (\emph{i.e.}, bag-level) $Y$ is available. Based on the MIL assumption, a positive bag should contain at least one positive instance, while a negative bag contains all negative instances \cite{andrews2003support, dietterich1997solving, amores2013multiple, cheplygina2019not} in a binary classification scenario. We build our system upon a bag-level MIL model with a parameterized attention module that aggregates instance features and forms the bag-level representation, instead of using a pre-defined function, such as maximum or mean pooling \cite{ilse2018attention}. Fig \ref{model_overview}. shows the overview of our model.

\label{subsec:attention_mil}
\subsection{Attention-based MIL with Instance Dropout}  
In the attention-based MIL model, a CNN is utilized to transform each instance into a $d$ dimensional feature vector $\mathbf{v}_i \in \mathbb{R}^{d}$. A permutation invariant function $f(\cdot)$ can be applied to aggregate and project $k$ instance-level feature vectors into a joint bag-level representation. We use a multilayer perceptron-based attention module as $f(\cdot)$ \cite{ilse2018attention}, which produces a combined bag-level feature vector $\mathbf{v'}$ and a set of attention values representing the relative contribution of each instance as defined in Eq \ref{eq:attention}.
\begin{align}\label{eq:attention}
\begin{split}
\mathbf{v'} &= f(\mathbf{V}) = \sum_{i=1}^{k} \alpha_i \mathbf{v}_i \\
\alpha &= \textrm{Softmax}[\mathbf{u}^T\mathrm{tanh}(\mathbf{W}\mathbf{V}^T)]
\end{split}
\end{align}
where $\mathbf{V} \in \mathbb{R}^{k \times d}$ contains the feature vectors for $k$ tiles, $\boldsymbol{u} \in \mathbb{R}^{h \times 1}$ and $\boldsymbol{W} \in \mathbb{R}^{h \times d}$ are parameters in the attention module, and $h$ denotes the dimension of the hidden layer. The slide-level prediction can be obtained by applying a fully connected layer to the bag-level representations $\mathbf{v'}$. Both the CNN feature extractor and the attention-based aggregation function are differentiable and can be trained end-to-end using gradient descent. The attention module not only provides a more flexible way to incorporate information from instances, but also enables us to localize informative tiles.  
However, this framework encounters similar problems as other saliency detection models \cite{zhang2018adversarial, hou2018self, singh2017hide}. In particular, as pointed out in \cite{ilse2018attention}, instead of detecting the all informative regions, the learned attention map can be highly sparse with very few positive instances having large values. This issue may be caused by the underlying MIL assumption that only one positive instance needs to be detected for a positive bag. This can affect the performance of our classification stage model, which relies on informative tiles selected by the learned attention map. To encourage the model to select more relevant tiles, we used an instance dropout method similar to \cite{singh2017hide, singh2018hide}. Specifically, instances are randomly dropped during the training, while all instances are used during model evaluation. To ensure the distribution of inputs for each node in the network remains the same during training and testing, pixel values of dropped instances are set to be the mean RGB value of the dataset \cite{singh2017hide, singh2018hide}. This form of instance dropout can be considered a regularization method that prevents the network from relying on only a few instances for bag-level classification. 

\begin{table*}[htbp]
\renewcommand{\arraystretch}{1.0}
  \centering
  \caption{Number of slides for each Grade group}
    \begin{tabular}{llllllll}
    \toprule
          & No. BN Slides & No. GG 1 Slides & No. GG 2 Slides & No. GG 3 Slides & No. GG 4 Slides & No. GG 5 Slides & No. Patients \\
    \midrule
    Train & 3,225  & 3,224  & 1,966  & 648   & 306   & 269   & 575 \\
    Validation & 2,579  & 412   & 307   & 95    & 17    & 67    & 86 \\
    Test  & 5,355  & 807   & 587   & 148   & 129   & 88    & 169 \\
    \midrule
    Total & 11,159 & 4,443  & 2,860  & 891   & 452   & 424   & 830 \\
    \bottomrule
    \end{tabular}%
  \label{tab:dataset}
\end{table*}%

\subsection{Attention-based Tile Selection}
An intuitive approach to localize suspicious regions with learned attention maps is to use the top $q$ percent of tiles with the highest attention weights. However, the percentage of cancerous regions can vary across different cases. Therefore, using a fixed $q$ may cause over selection for slides with small suspicious regions and under selection for those with large suspicious regions. Moreover, this method relies on an attention map, which in this context is learned without explicit supervision at the pixel- or region-level. To address these challenges, we incorporate information embedded in instance-level representations by selecting informative tiles from clusters. Specifically, instance representations obtained from the MIL model are projected to a compact latent embedding space using principle component analysis (PCA). We then perform K-means clustering to group instances with similar semantic features based on their PCA transformed instance-level representations. The relevance of each cluster $\bar{\alpha}_{s}$ can be determined by the average attention weights of tiles within it as defined by $\bar{\alpha}_{s} = \frac{1}{m}\sum^{m}_{j=1}{\alpha}_j$. The intuition is that clusters that contain more relevant information for slide classification should have higher average attention weights. For example, in a cancer-positive slide, clusters consisting of cancerous glands should have higher attention weights compared to those with benign glands and stromal regions. Finally, we can determine the number of tiles to extract from each cluster based on the $\bar{\alpha}_{s}$ and the total number of tiles.  

\subsection{Multi-resolution WSI classification}
Different from most medical imaging modalities, WSIs typically contain billions of pixels, which make them practically impossible to feed into GPU memory directly at full resolution. Though the size of WSIs is enormous, most regions typically do not contain relevant information for slide classification, such as stroma and benign glands. Pathologists tend to analyze the entire slide at a relatively low resolution, usually at 5x, to find suspicious regions and then switch to higher magnification in these areas to render a final diagnosis. Our proposed MRMIL model mimics this process, containing two stages as shown in Fig \ref{model_overview}. The detection stage model, which consists of an attention-based MIL with instance dropout, is trained with all tiles extracted at a lower magnification (\emph{i.e.}, at 5x) to differentiate benign and malignant slides and generate attention maps. The attention-based clustering method is applied to select relevant tiles for the classification stage model. Selected tiles are extracted at the same location, but at a higher magnification (\emph{i.e.} at 10x) and fed into the MIL network for cancer grade prediction. 

%% file: tex/experiment.tex
\label{sec:experiment}
\subsection{Dataset and data preprocessing}
\label{sec:dataset}
\subsubsection{Dataset} Our dataset contains 20,229 slides from prostate needle bihttps://www.overleaf.com/project/5e9ddafd1787bf0001e79fc9opsies from 830 patients pre- or post-diagnosis (IRB16-001361). Slides' labels extracted from their corresponding pathology reports. There are no additional fine-grained annotations at the pixel- or region-level for this dataset. Additionally, we did not rely on any pre-trained tissue, epithelium, or cancer segmentation networks, and did not perform extensive manual curation to exclude slides with artifacts such as air bubbles, pen markers, dust, \emph{etc}. We randomly divided the dataset into 70\% for training, 10\% for validation, and 20\% for testing, stratifying by patient-level GG determined by the highest GG in each patient's set of biopsy cores. This process produced a test set with 7,114 slides from 169 patients and a validation set containing 3,477 slides from 86 patients. From the rest of the dataset, we balanced sampled benign (BN), low grade (LG), and high grade (HG) slides. Table \ref{tab:dataset} shows more details on the breakdown of slides.
\subsubsection{External dataset} We evaluated our models on a public prostate dataset, SICAPV1, collected by the Hospital Clínico Universitario de Valencia, which contains $512 \times 512$ tiles at 10x extracted from 79 slides of prostate needle biopsies with 50\% overlapping \cite{Esteban2019}. 19 of these slides are benign, and the rest are malignant. 

\subsubsection{Data preprocessing} The majority of regions on WSIs are background. Thus, we converted slides downsampled at their lowest available magnification compressed in the .svs file into HSV color space and thresholded on the hue channel to produce tissue masks. Morphological operations such as dilation and erosion were used to fill in small gaps, remove isolated points, and further refine tissue masks. We then extracted tiles of size $256 \times 256$ at 10x from the grid with 12.5\% overlap. Tiles that contain less than 80\% tissue were discarded from analysis. The number of tiles per slide ranges from 1 to 1,273, with an average of 275. To account for stain variability, we used a color transfer method \cite{reinhard2001color} to normalize tiles extracted from the slide. The scanning objective was set at 20x (0.5 $\mu$m per pixel). We downsampled tiles to 5x for the detection stage model development.  
For external dataset, we divided the $512 \times 512$ tiles into 4 non-overlapping $256 \times 256$ sub-tiles, in order to match the input size of our models. The same stain normalization \cite{reinhard2001color} was applied.

\begin{table*}[!htbp]
  \centering
  \caption{Model performance on BN, LG, and HG slides classification}
    \begin{tabular}{lrllllll}
    \toprule
          &       &       & \multicolumn{2}{c}{BN, LG, HG Classification} &       & \multicolumn{2}{c}{Cancer Detection} \\
\cmidrule{4-5}\cmidrule{7-8}    Experiment Name &       & Model Details & Cohen's Kappa (\%) & Acc (\%) &       & AUROC (\%) & AP (\%) \\
    \midrule
    Handcrafted + RF &       & 90 radiomics features + RF at 10x & 56.95 & 81.5  &       & 93.1  & 83.9 \\
    Handcrafted + Xgboost &  &  90 radiomics features + Xgboost at 10x  & 55.92 & 80.9  &       & 93.3  & 83.9 \\
    \midrule
    Campanella \emph{et-al.} \cite{campanella2019clinical} &       & MIL + RNN at 10x & 77.2  & 90.7  &       & 98.3  & 97.3 \\
    Mean aggregation &       & Mean aggregation at 10x & 77.1  & 90.8  &       & 97.9  & 96.6 \\
    Max aggregation &       & Max aggreagtion at 10x & 79.5  & 91.9  &       & 97.4  & 96.3 \\
    \midrule
    Single stage &       & Single resolution MIL at 5x & 76.3  & 92.5  &       & 97.4  & 95.8 \\
    Br selection &       & Multi-resolution + Br   & 76.0  & 90.8  &       & 95.9  & 94.3 \\
    W/o instance dropout &       & Multi-resolution + Att & 77.3  & 91.0  &       & 97.3  & 96.0 \\
    Att selection &       & Multi-resolution + Att + instance dropout & 80.7  & 92.4  &       & \textbf{98.4} & \textbf{97.4} \\
    MRMIL &       & Multi-resolution + Att + instance dropout + clusters & \textbf{81.8} & \textbf{92.7} &       & 98.2  & \textbf{97.4} \\
    \bottomrule
    \end{tabular}%
  \label{tab:result_summary}%
\end{table*}%

\subsection{Implementation Details}
\label{subsec:exp:implementation}
We used VGG11 with batch normalization (VGG11bn) \cite{simonyan2014very} as the backbone for the feature extractor in the MRMIL model for both detection stage and classification stage. A $1 \times 1$ convolutional layer was added after the last convolutional layer of VGG11bn to reduce dimensionality and generate $k \times 256 \times 4 \times 4 $ instance-level feature maps for $k$ tiles. Feature maps were flattened and fed into a fully connected layer with 256 nodes, followed by ReLU and dropout layers. This produced a $k \times 256$ instance embedding matrix, which was forwarded into the the attention module. The attention part, which generated a $k \times n$ attention matrix for $n$ prediction classes, consisted of two fully connected layers with dropout, tanh non-linear activations, and a softmax layer. Instance embeddings were multiplied with attention weights, resulting in a $n \times 256$ bag-level representation, which was flattened and input into the final classifier. The probability of instance dropout was set to $0.5$ for both model stages.

The CNN feature extractor was initialized with weights learned from the ImageNet dataset \cite{deng2009imagenet}. After training the attention module and the classifier with the feature extractor frozen for three epochs, we trained the last three VGG blocks together with the attention module and classifier for 97 epochs. The initial learning rates for the feature extractor were set at $1 \times 10^{-5}$ and $5 \times 10^{-5}$ for the attention module and the classifier, respectively. The learning rate was decreased by a factor of 10 if the validation loss did not improve for the last 10 epochs. We used the Adam optimizer \cite{kingma2014adam} and a batch size of one. Detection stage and classification stage models were trained separately using the same training hyperparameter (\emph{e.g.,} learning rate, batch size and \emph{etc.}.).

For clustering-based region selection, we projected $k \times 256$ instance embedding matrix to $k \times 32$ with PCA, and utilized K-Means clustering to group tiles. The number of clusters was set to be 3 to encourage tiles to be grouped into LG, HG and BN clusters.   

Hyper-parameters were tuned on the validation set.
We further extended our MRMIL model for GG prediction. The cross entropy loss weighted by reversed class frequency was utilized to address the class imbalance problem.
Hyperparameters were selected using the validation set.   
Models were implemented in PyTorch 0.4.1 \cite{paszke2017automatic}, and trained on an NVIDIA DGX-1.

\subsection{Evaluation Metrics}  
As our test dataset contained over 75\% benign slides, accuracy (Acc) alone is biased metric for model evaluation. In addition, we used the AUROC and AP computed from ROC and precision and recall (PR) curves, respectively. For cancer grade classification, we measured the Cohen's Kappa ($\kappa$), $\kappa = \frac{p_o - p_e}{1 - p_e}$. $p_o$ is the agreement between observers and $p_e$ is the probability of agreement by chance. 
All metrics were computed using the scikit-learn 0.20.0 package \cite{scikit-learn}.  

\subsection{Model Visualization}
In addition to quantitative evaluation metrics, interpretability is important in developing explainable machine learning tools, especially for medical applications. In order to have a better understanding of our model predictions, we performed t-Distributed Stochastic Neighbor Embedding (t-SNE) \cite{maaten2008visualizing} of learned bag-level representations for both stage models. Specifically, for each slide we utilized the flattened $n \times 256$ feature vector before being forwarded to the final classification layer. The learning rate of t-SNE was set at $1.5 \times 10^2$, and the perplexity was set at $30$.  

The saliency map produced by the attention module in the MRMIL model only demonstrated the relative importance of each tile. To further localize discriminative regions within tiles, we utilized Gradient-weighted Class Activation Mapping (Grad-CAM) \cite{selvaraju2017grad}. Concretely, given a trained MRMIL model and a target class $c$, we retreived the top $k$ tiles with the highest attention weights, which were fed to the model to compute gradients and activations. 

\subsection{Model Comparison}
{\noindent \bf Handcrafted features}. We converted input tiles at 10x into HSV color space and thresholded on the H channel to get tissue masks. Then we utilized the PyRadiomics package \cite{van2017computational} to extract 90 features for each tile, including 16 first-order statistics, 23 gray level co-occurrence matrix-based, 16 gray level run length matrix-based, 16 gray level size zone matrix-based, 5 neighbouring gray tone difference matrix-based, and 14 gray level dependence matrix-based features. The maximum pooling was applied to aggregate tile-level features, which were fed into the final slide-level classifier. We experimented with Xgboost \cite{chen2016xgboost} and random forest (RF) \cite{liaw2002classification} classifiers. Grid search with 3-fold cross validation was used to select hyperparameters for classifiers.   

{\noindent \bf MIL model by Campanella \emph{et-al.} \cite{campanella2019clinical}}. We compared our model with the related recent work \cite{campanella2019clinical}, which also trained slide classification models with only slide-level labels in the MIL framework. Different from our model, they utilized an instance-level MIL approach. The CNN model was trained on top k tiles with high probability after applying the partially trained model, and this process is iterated for certain epochs. Then they utilized the RNN model to aggregate features from top k tiles for final classification. We used the implementation provided by \cite{campanella2019clinical} and hyperparameters reported in the paper to re-train the model on our dataset.   

{\noindent \bf Different aggregation methods}. Instead of using the attention module to aggregate tile-level features to slide-level representations, we experimented with different aggregation methods: maximum pooling and mean pooling aggregation. 

{\noindent \bf Single stage}. To evaluate the effectiveness of the multi-
resolution model in cancer grade prediction, we compared our model with a model trained with all extracted tiles at 5x only, referred as Single stage.  

{\noindent \bf Blue ratio selection}. Blue ratio (Br) image conversion, as defined in Eq \ref{eq:blue_ratio}, can accentuate the blue channel of a RGB image and thus highlight proliferate nuclei regions \cite{chang2012nuclear}. 

\begin{align} \label{eq:blue_ratio}
\textrm{Br} = \frac{100 \times B}{1 + R + G} \times \frac{256}{1 + R + G + B}
\end{align}
where $R$, $G$, $B$ are the red, green and blue channels in the original RGB image.   
Br conversion is one of the most commonly used approaches to detect nuclei 
\cite{chang2012nuclear,saha2018efficient} and select informative regions from large-scale WSIs \cite{del2017convolutional, arvaniti2018coupling, lawson2019persistent}. To evaluate the attention-based ROI detection, we replaced the first stage cancer detection model with the Br conversion to select the top $q = 25\%$ tiles with highest average Br values, referred to as \textit{br selection}.   

{\noindent \bf Without instance dropout}. In this experiment, denoted as \textit{w/o instance dropout}, we investigated whether instance dropout could improve the integrity of learned attention map and lead to better performance.  

{\noindent \bf Attention-only selection}. Instead of selecting informative clusters, we only utilized the attention map by choosing the top $q = 25\%$ tiles with the highest attention values as the input for the second stage model in the \textit{att selection} experiment. 

%% file: tex/result.tex
\begin{figure}
\includegraphics[width=3.5in]{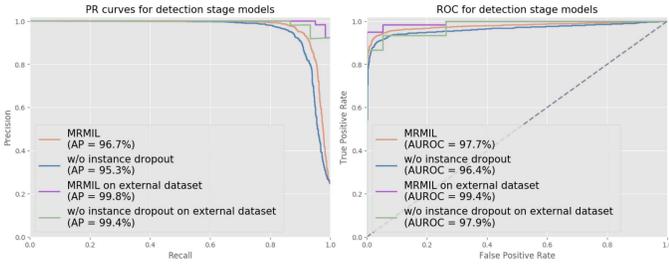}
\caption{ROC and PR curves for detection stage models on our test set and external dataset. In the detection stage, models were trained to distinguish malignant and benign slides with all tiles extracted from slides at 5x.}
\label{fig:pr_roc_stage1}
\end{figure}

\begin{figure}
\includegraphics[width=2.8in]{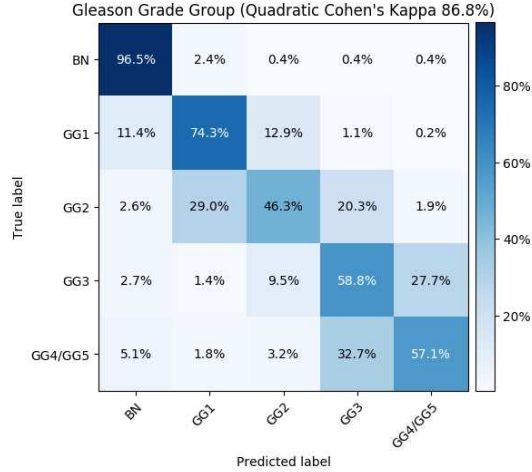}
\caption{Confusion matrix for GG prediction.}
\label{fig:cm_stage2}
\end{figure}

\begin{figure}
\centering
\includegraphics[width=3.4in]{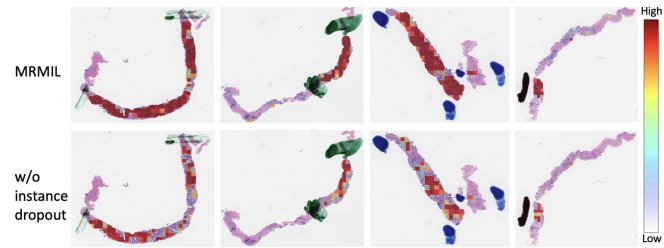}
\caption{WSIs overlaid with attention maps generated from the first stage cancer detection model. Pen markers as mentioned in Section \ref{sec:dataset} indicate cancerous regions. The first row shows attention maps from the model with instance dropout, while the second row is from the model without using instance dropout. Figures are best viewed in color.}
\label{fig:attention_maps}
\end{figure}

\begin{figure}
\centering
\includegraphics[width=3.5in]{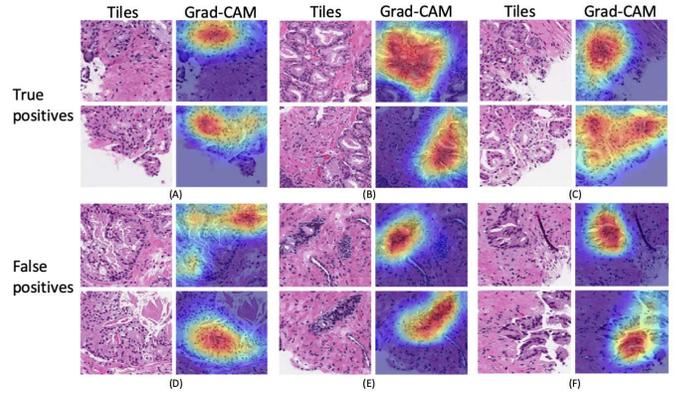}
\caption{Visualization of discriminative regions within tiles for TP and FP predictions. For each slide (A)-(F), we selected the top two tiles with the highest attention weights from the model, which were then forwarded to the model to generate activations and gradients for Grad-CAM.}

\label{fig:grad_cam}
\end{figure} 

\begin{figure}
\centering
\includegraphics[width=3.4in]{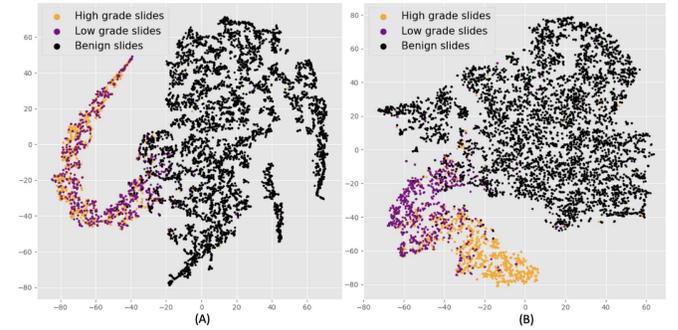}
\caption{t-SNE visualization of slide-level features. Black dots denote benign, purple dots indicate LG, and orange dots represent HG slides. (A) is the slide-level representations from the detection stage model. There is distinct separation between benign and cancerous slides. (B) shows the slide-level features from the classification stage model. We can see a better separation between LG and HG slides.}

\label{fig:tsne_plots}
\end{figure}

Fig \ref{fig:pr_roc_stage1} shows both ROC and PR curves for the detection stage cancer models trained at 5x. The detection stage model in the MRMIL obtained an AUROC of 97.7\% and an AP of 96.7\% on our internal test set. On the external dataset, it achieved an AUROC of 99.4\% and an AP of 99.8\%.  The model trained without using the instance dropout method yielded a slightly lower AUROC and AP on both internal and external datasets. 

Since our dataset does not have fine-grained annotations, we visualized generated attention maps and compared them with pen markers annotated by pathologists during diagnosis. We masked out markers as mentioned in Section \ref{sec:dataset}, thus they were not utilized for model training. Fig \ref{fig:attention_maps} presents the comparison between attention maps learned from models with and without using instance dropout during training.  

To further localize suspicious regions within a tile and better interpret model predictions, we applied Grad-CAM on the first detection stage MIL model as shown in Fig \ref{fig:grad_cam}. We generated Grad-CAM maps for not only true positives (TP), but also false positives (FP) to understand which parts of the tile led to false predictions. We selected two tiles with highest attention weights from each slide for visualization.

The MRMIL model projects input tiles to embedding vectors, which are aggregated and form slide-level representations. The t-SNE method enables high dimensional slide-level features to be visualized at a two dimensional space as demonstrated in Fig \ref{fig:tsne_plots}. Fig \ref{fig:tsne_plots} (A) is the t-SNE plot for the detection stage model and (B) presents bag-level features produced by the classification stage model with selected high resolution tiles as inputs. 

Table \ref{tab:result_summary} shows model performances on BN, LG, HG classification. The proposed MRMIL outperformed all baseline models and achieved the highest Acc of 92.7\% and $\kappa$ of 81.8\% as shown in row 12. Models with handcrafted features only obtained about 57\% $\kappa$ as demonstrated in row 3 and 4 in the Table \ref{tab:result_summary}. As shown in row 5, the model by Campanella \emph{et-al.} \cite{campanella2019clinical} got 4\% lower $\kappa$ compared with our MRMIL model. Models with simple mean and maximum pooling aggregation methods also achieved lower performance than the MRMIL model as reported in row 6 and 7. Row 8 to 11 demonstrated results on ablation study of the MRMIL model. The single stage attention MIL model trained at 5x achieved 76.3\% $\kappa$. The br selection that relied on the Br image for tile selection only obtained an Acc of 90.8\% and a $\kappa$ of 76.0\%. The w/o instance dropout model, got roughly 4\% lower $\kappa$ and 2\% lower Acc compared with the MRMIL model. In addition, we combined LG and HG predictions from the classification model and computed the AUROC and AP for detecting cancerous slides. For instance, by zooming in on suspicious regions identified by the detection stage model, the MRMIL achieved an AUROC of 98.2\% and an AP of 97.4\%, both of which are higher than the detection stage only model. We present the confusion matrix for the MRMIL model on GG prediction in Fig \ref{fig:cm_stage2}. The MRMIL model obtained an accuracy of 87.9\%, a quadratic $\kappa$ of 86.8\%, and a $\kappa$ of 71.1\% for GG prediction.

%% file: tex/discussion.tex
\label{sec:discussion}
Our detection stage model achieved promising results on both an internal test set and an external dataset, which demonstrates the generalizability of the model. One potential explanation for slightly better performances on external dataset is that our independent test set is relatively large (\emph{i.e.} 7114 slides from 830 patients.) and is collected from clinical database without any data curation.   

Handcrafted features-based models performed relatively well on differentiating benign and malignant slide with an AUC of 93.3\%, however, they obtained much lower $\kappa$ on the hard task of classifying LG, HG and BN slides. The model proposed by Campanella \emph{et-al.} \cite{campanella2019clinical} first used an instance-based MIL approach, which considered tiles with highest probabilities as having the same label as the corresponding slide, and then utilized the RNN model to aggregate representations from top tiles for slide classification. In contrast, our model used a more flexible attention aggregation method that can detect discriminative tiles and combine tile-level features in the same time. The model \cite{campanella2019clinical} achieved comparable performance on detecting cancerous slides with 98.3\% AUC and 97.3\% AP. Yet, it showed inferior results on predicting LG, HG, and BN classes compared with the MRMIL model.   

The quality of attention maps from the detection stage model is essential for selecting discriminative regions for the classification stage model. As shown in Fig \ref{fig:attention_maps}, attention maps learned with only weak (\emph{i.e.} slide-level) labels are consistent with cancerous regions identified by pathologists during diagnosis. This demonstrates that our detection stage model not only achieves strong performance in classifying malignant versus benign slides, but also identifies suspicious regions for classification stage models. In addition, the generated attention maps can be integrated into a WSI viewer to potentially help pathologists more quickly localize relevant areas and reduce diagnostic time. Fig \ref{fig:attention_maps} also shows that the original attention-based MIL model \cite{ilse2018attention} (\emph{i.e.} w/o instance dropout) only focuses on a few most discriminative tiles instead of entire suspicious regions. As reflected in Table \ref{tab:result_summary}, the w/o instance dropout model obtained a $\kappa$ of 77.3\%, which is about 4\% lower than the one trained with instance dropout. Moreover, the performance of the model that relied on the Br image is inferior to the models that utilized attention maps. This demonstrates that areas with the most blue color may not be diagnostic relevant regions and that our attention module is able to extract high-level predictive representations rather than purely color features.  

Grad-CAM visualization facilitates understanding of predictions from 'black-box' deep learning models, as shown in Fig \ref{fig:grad_cam}. For TP predictions in Fig \ref{fig:grad_cam} (A)-(C), our model captured the most relevant parts in the tile, though some cancerous regions were missed. For example, the first tile in (B) contains densely clustered cancerous glands, but the corresponding Grad-CAM only highlighted the most central area, and cancerous glands closer to the boundary were not detected. FP predictions are usually also hard cases for pathologists, with features that resemble prostate cancer. For example, regions highlighted by Grad-CAM in (E) contain benign glands with increased number of basal cells due to tangential tissue sectioning. (F) in Fig \ref{fig:grad_cam} shows the seminal vesicle/ejaculatory duct tissue that form small outpouching glands with amphophilic cytoplasm, which mimic malignant glands. Our model was only trained to detect and grade acinar adenocarcinoma for prostate biopsies. Interestingly, as shown in (D), the model was able to identify intraductal carcinoma of the prostate gland (IDC-P), which is usually associated with high-stage invasive cancer and adverse prognosis.

From Fig \ref{fig:tsne_plots} (A), we can see that benign slide representations are clustered together on the right and malignant slides form a small cluster on the left. There is no distinct separation between features from LG and HG slides, since the objective of the detection stage model is to classify cancerous versus benign slides. Fig \ref{fig:tsne_plots} (B) shows that features of LG and HG slides generated from the classification stage model form their own distinct clusters, and representations from LG slides lie closer to benign slides in the embedding space.  

To quantitatively evaluate our model performance, we performed experiments to understand the contribution of different model components, as summarized in Table \ref{tab:result_summary}. Using attention maps to select higher resolution tiles improved the $\kappa$ of the one with br selection by 1\%. Instance dropout further boosted the $\kappa$ by over 3\%. The final model MRMIL with all components achieved the highest $\kappa$ for BN, LG, and HG classification, 98.2\% AUROC for detecting malignant slides, and a quadratic $\kappa$ of 86.8\% for GG prediction, which is comparable to state-of-the-art models that require pre-trained segmentation networks \cite{bulten2019automated}.

%% file: tex/limitations.tex
In this section, we discuss limitations of this work and some potential directions for future research.
In this work, we developed a two-stage model to first detect suspicious regions and then classify cancer grade with selected tiles at a higher magnification. The tile selection is determined by the detection stage model, and there is no mechanism to adaptively update selected tiles according to the loss from the classification stage model. In future work, a recurrent network or reinforcement learning can be incorporated to dynamically resample suspicious regions during training.  

We only developed the model for acinar adenocarcinoma detection and classification for prostate biopsies. Other prognostically relevant histopathological types, such as ductal adenocarcinoma and IDC-P, need to be investigated in future studies. 

In this study, we merely qualitatively evaluated our attention maps by visually inspecting learned maps for slides with pen markers. Though our model was able to identify similar regions as indicated by pen markers, quantitative evaluation with manual region-level annotations could provide a better metric for the attention module.   

Additionally, besides achieving promising $\kappa$ and Acc, a successful CAD tool should be able to facilitate clinical diagnosis. In future work, we will investigate different approaches to evaluate the effectiveness of our model as a CAD tool.

%% file: tex/conclusion.tex
In this paper, we developed a novel MRMIL model that consists of a detection stage and a grade classification stage. The model can be trained with weak supervision from slide-level labels and localize cancerous regions. We provided visualization of saliency maps at both the slide- and tile-level, and learned representations to enhance model interpretability. The model was developed and evaluated on a dataset with over 20k prostate slides from 830 patients and an external dataset \cite{Esteban2019}, and achieved promising performance. We believe that these types of models could have multiple applications in the clinic, including allowing pathologists to increase their efficiency, empowering more general pathologists to perform at the level of experts, and performing "second reads" of biopsy slides for quality assurance.